\documentstyle[amssymb,prl,aps,epsfig,floats,twocolumn]{revtex}
\begin{document}
%\draft

\wideabs{

\title{Spin Waves and Electronic Interactions in La$_2$CuO$_4$}
\author{R. Coldea$^{1,2}$, S.M. Hayden$^3$,
G. Aeppli$^4$, T.G. Perring$^2$, C.D. Frost$^2$,
T.E. Mason$^1$, S.-W. Cheong$^5$, and Z. Fisk$^6$}
\address{
$^1$Oak Ridge National Laboratory, Oak Ridge, TN 37831, USA\\
$^2$ISIS Facility, Rutherford Appleton Laboratory, Chilton,
Didcot, OX11 0QX, UK\\
$^3$H.H. Wills Physics Laboratory, University of Bristol, Bristol,
BS8 1TL, UK\\
$^4$NEC Research Institute, Princeton, NJ 08540, USA \\
$^5$Lucent Technologies, Murray Hill, NJ 07974, USA \\
$^6$Department of Physics, Florida State University, Tallahassee,
Florida 32306, USA}

\date{5 April 2001}

\maketitle

\begin{abstract}
The magnetic excitations of the square-lattice spin-1/2 antiferromagnet
and high-Tc parent La$_2$CuO$_4$ are determined using high-resolution inelastic
neutron scattering. Sharp spin waves with absolute intensities in agreement
with theory including quantum corrections are found throughout the Brillouin
zone. The observed dispersion relation shows evidence for substantial
interactions beyond the nearest-neighbor Heisenberg term, which can
be understood in terms of a cyclic or ring exchange due to the strong
hybridization path around the Cu$_4$O$_4$ square plaquettes.
\end{abstract}

  % close \pacs
}  % close \wideabs

\narrowtext

\begin{figure}[t]
\centering
\epsfysize=11cm
%\vspace{11cm}
\epsffile{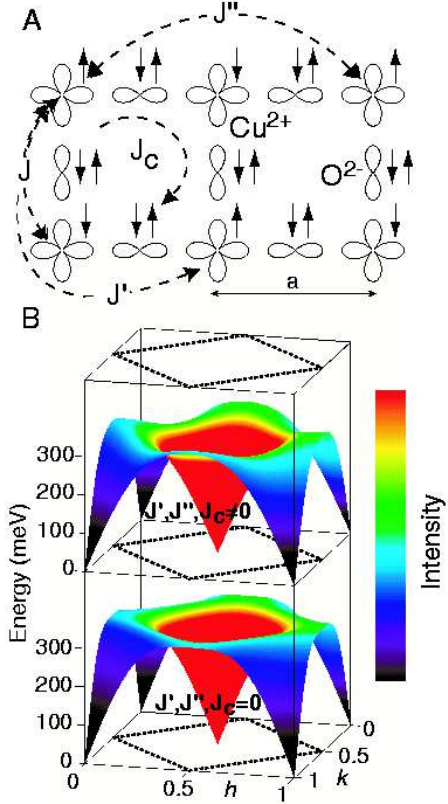}
\caption{ (color) {\bf A} The
CuO$_2$ plane showing the atomic orbitals (Cu 3$d_{x^2-y^2}$ and O
2$p_{x,y}$) involved in the magnetic interactions. $J$,
$J^{\prime}$ and $J^{\prime\prime}$ are the first-, second- and
third-nearest-neighbor exchanges and $J_c$ is the cyclic
interaction which couples spins at the corners of a square
plaquette. Arrows indicate the spins of the valence electrons
involved in the exchange. {\bf B} Lower surface is the dispersion
relation for $J$=136 meV and no higher-order magnetic couplings or
quantum corrections. The upper surface shows the effect of the
higher-order magnetic interactions determined by the present
experiment. Color is spin-wave intensity.} \label{fig1}
\end{figure}

While there is consensus about the basic phenomenology - electron
pairs with non-zero angular momentum, unconventional metallic
behavior in the normal state, tendencies towards inhomogeneous
charge and spin density order - of the high temperature copper
oxide superconductors, there is no agreement about the microscopic
mechanism. After over a decade of intense activity, there is not
even consensus as to the simplest ``effective Hamiltonian'', which
is a short-hand description of the motions and interactions of the
valence electrons, needed to account for cuprate
superconductivity. Because much speculation is centered on
magnetic mechanisms for the superconductivity, it is important to
identify the interactions among the spins derived from the
unfilled Cu$^{2+}$ $d$-shells. The present experiments show that
there are significant (on the scale of the pairing energies for
high-Tc superconductivity) interactions coupling spins at
distances beyond the 3.8 \AA\ separation of nearest-neighbor
Cu$^{2+}$ ions. Cyclic or ring exchange due to a strong
hybridization path around the Cu$_4$O$_4$ squares (see Fig.\
\ref{fig1}A), from which the cuprates are built, provides a
natural explanation for the measured dispersion relation.
CuO$_2$ planes are thus the second example %\cite{Furrer92}
of an important Fermi system ($^3$He is the other \cite{Roger83})
where significant cyclic exchange terms have been deduced.

Magnetic interactions are revealed through the wavevector dependence or dispersion of the
magnetic excitations. In magnetically ordered materials, the dominant excitations are spin
waves which are coherent (from site to site as well as in time) precessions of the spins about
their mean values. The lower frame of Fig.\ \ref{fig1}B shows the dispersion relation calculated
using conventional linear spin-wave theory in the classical large-$S$ limit, where the only
magnetic interaction is a strong nearest-neighbor superexchange coupling $J$ \cite{Anderson63}.
We identify wavevectors by their coordinates ($h,k$) in the two-dimensional (2D) reciprocal
space of the square lattice. Spin waves emerge from the wavevector (1/2,1/2) characterizing
the simple antiferromagnetic (AF) unit cell doubling in La$_2$CuO$_4$ \cite{Vaknin87},
and disperse to reach a maximum energy $2J$ that is a constant along the AF zone boundary
marked by dashed squares in Fig.\ \ref{fig1}B. Longer-range interactions manifest themselves most
simply at the zone boundary. The upper frame of Fig.\ \ref{fig1}B shows the dispersion calculated
with modest interactions between next nearest-neighbors. Virtually the only visible effect
of the additional interactions is the dispersion of the spin waves along the zone edge. Thus,
experiments to test for such interactions must measure the spin waves along the zone boundary.
Only inelastic neutron scattering with high energy and wavevector resolution can accomplish
this, although photon spectroscopy \cite{Lyons89,Sugai90,Lorenzana99,Perkins93} has led to
suspicions of such interactions.

For La$_2$CuO$_4$, a requirement that complicates meeting the
resolution goals is the need to use neutrons with energies in the
epithermal, 0.1-1.0 eV, range rather than in the more conventional
cold and thermal \cite{Shirane87}, 2-50 meV, regimes. An early
high energy neutron scattering experiment \cite{Hayden91} revealed
well-defined spin-wave excitations throughout the Brillouin zone
which could be modeled using a nearest-neighbor Heisenberg
exchange $J$=136 meV. The directions of the scattered neutrons
were specified only to within the solid angle determined by the
large detector dimensions. Thus, the measured spectra represented
averages over large portions of the reciprocal space, so that
dispersion along the zone boundary was unresolvable and only an
upper bound could be placed on further neighbor couplings. The
advance enabling the present investigation is the use of
position-sensitive detectors for the scattered neutrons, which
increases the wavevector resolution by an order of magnitude. The
new detector bank is installed in the direct-geometry High-Energy
Transfer (HET) time-of-flight spectrometer at the ISIS
proton-driven pulsed neutron spallation source.

\begin{figure}[t]
\centering
\epsfxsize=8.0cm
\epsffile{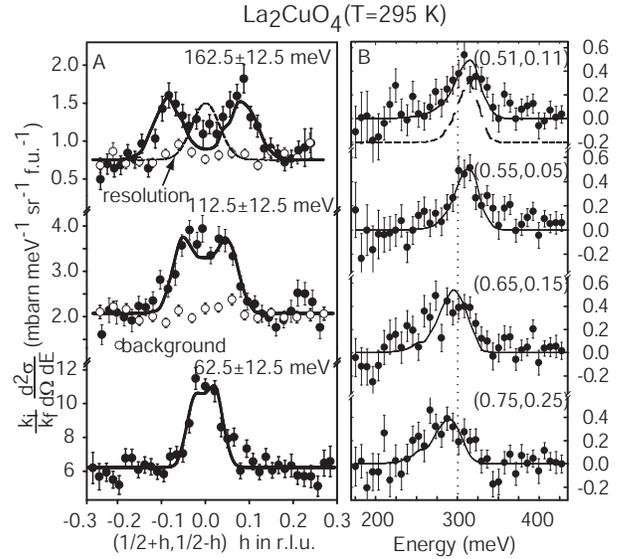} \vspace{0.5cm}
\caption{ Scattering from the spin waves in La$_2$CuO$_4$ ($T$=295
K). Data result from 68 hours (A) and 98 hours (B) counting at a
proton current of 170 $\mu$A. The sample is described in
\protect\cite{sample_prep}. Solid lines are fits to a spin-wave
cross-section convolved with the instrumental resolution. (A)
Const-$E$ cuts near the AF zone center for an incident energy
$E_i$=250 meV. $Q_z$ wavevector components at scan centers are
$l$=2.8 (bottom panel), 6.2 and 9.6 r.l.u. Open circles are a
background measured near the (0,0) position. Dashed curve is the
instrumental response to spin waves of infinite velocity. (B)
Const-$Q$ cuts, with $E_i$=750 meV, yield the dispersion along the
AF zone boundary. Vertical dotted line at $E$=300 meV is a guide
to the eye. $l$ values at peak position vary from 8.8 (bottom
panel) to 9.5 (top panel). A background measured near the nuclear
zone center (1,0) has been subtracted. Dashed curve is the
instrumental response to a dispersionless mode.}
%As in other quasi-2D antiferromagnets, damping effects are expected to be
%small at temperatures $k_BT \ll J$ \protect\cite{Kopietz90} and therefore
%well-defined excitations can still be observed throughout the Brillouin
%zone for energies $E \gg k_BT$.}
\label{fig2}
\end{figure}

\begin{figure}[t]
\centering \epsfxsize=7.0cm
\epsffile{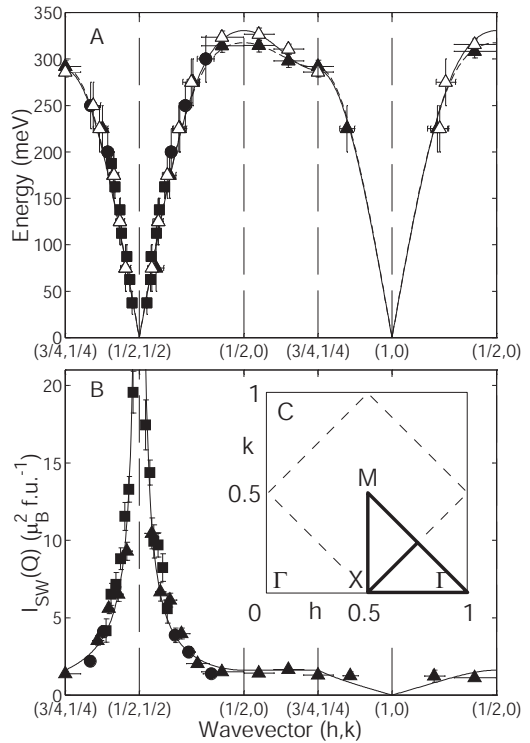} \caption{
(A) Dispersion relation along high symmetry directions in the 2D
Brillouin zone, see inset (C), at $T$=10 K (open symbols) and 295
K (solid symbols). Squares were obtained for $E_i$=250 meV,
circles for $E_i$=600 meV and triangles for $E_i$=750 meV. Points
extracted from constant-$E$(-$Q$) cuts have a vertical
(horizontal) bar to indicate the $E$($Q$) integration band. Solid
(dashed) line is a fit to the spin-wave dispersion relation at
$T$=10 K (295 K) as discussed in the text. (B)
Wavevector-dependence of the spin-wave intensity at $T$=295 K
compared with predictions of linear spin-wave theory shown by the
solid line. The absolute intensities \protect\cite{intensities}
yield a wavevector-independent intensity-lowering renormalization
factor of 0.51$\pm$0.13 in agreement with the theoretical
prediction of 0.61 \protect\cite{Singh89} that includes the
effects of quantum fluctuations.} \label{fig3}
\end{figure}

Fig.\ \ref{fig2}A shows data in the form of constant energy scans for wavevectors
around the antiferromagnetic zone center. As $E$ increases, counter-propagating modes
become apparent.
%%%%%%%%%%%%%%%%%%%%%%%%%%%%%%%%%%%%%%%%%%%%%%%%%%%%%%%%%%%%%%%%%%%%%%%%%%%%%%%
%%%%%%%%%%%%%%%%%%%%%%%%%%%%%%%%%%%%%%%%%%%%%%%%%%%%%%%%%%%%%%%%%%%%%%%%%%%%%%%
%%%%%%%%%%%%%%%%%%%%%%%%%%%%%%%%%%%%%%%%%%%%%%%%%%%%%%%%%%%%%%%%%%%%%%%%%%%%%%%%%%
%%%%%%%%%%%%%%%%%%%%%%%%%%%%%%%%%%%%%%%%%%%%%%%%%%%%%%%%%%%%%%%%%%%%%%%%%%%%%%%%%%
As the zone boundary is approached and there is less dispersion, inspection of
Fig.\ \ref{fig1}B reveals that it should be easier to locate the spin waves via energy
scans performed at fixed wavevector. Fig.\ \ref{fig2}B shows a series of such scans
collected at various points along the zone boundary. The spin waves have a clearly
noticeable dispersion, from a minimum of 292$\pm$7 meV near ${\bbox Q}$=(3/4,1/4)
to a maximum of 314$\pm$7 meV near (1/2,0). This is in obvious contrast
to the dispersion-less behavior of linear spin-wave theory for the nearest-neighbor
Heisenberg model. We have collected data throughout the Brillouin zone
%both at $T$=10 K and 295 K
and Fig.\ \ref{fig3}A shows the resulting dispersion along
major symmetry directions obtained from cuts of the type shown in Fig.\ \ref{fig2}.
Fig.\ \ref{fig3}B displays the corresponding spin-wave intensities,
in absolute units calibrated using acoustic phonon scattering from the sample.\\
%%%%%%%%%%%%%%%%%%%%%%%%%%%%%%%%%%%%%%%%%%%%%%%%%%%%%%%%%%%%%%%%%%%%%%%%%%%%%%%%%%
%%%%%%%%%%%%%%%%%%%%%%%%%%%%%%%%%%%%%%%%%%%%%%%%%%%%%%%%%%%%%%%%%%%%%%%%%%%%%%%%%%
%%%%%%%%%%%%%%%%%%%%%%%%%%%%%%%%%%%%%%%%%%%%%%%%%%%%%%%%%%%%%%%%%%%%%%%%%%%%%%%%%%
%%%%%%%%%%%%%%%%%%%%%%%%%%%%%%%%%%%%%%%%%%%%%%%%%%%%%%%%%%%%%%%%%%%%%%%%%%%%%%%%%%
To understand our results, we consider a Heisenberg Hamiltonian including higher
order couplings \cite{Takahashi77,Roger89,MacDonald90,Peters88}
%%%%%%%%%%%%%%%%%%%%%%%%%%%%%%%%%%%%%%%%%%%%%%%%%%%%%%%%%%%%%%%%%%%%%%%%%
\begin{eqnarray}
\label{eq_ham}
{\cal H} & = & J \sum_{\langle i,j \rangle} {\bbox S}_i \cdot {\bbox S}_j +
 J^{\prime} \sum_{\langle i,i^{\prime} \rangle} {\bbox S}_i \cdot {\bbox S}_{i^{\prime}} +
 J^{\prime\prime} \sum_{\langle i,i^{\prime\prime} \rangle} {\bbox S}_i \cdot {\bbox S}_{i^{\prime\prime}} \nonumber \\
& &
+ J_c \sum_{\langle i,j,k,l \rangle}
\left\{ ({\bbox S}_i \cdot {\bbox S}_j)({\bbox S}_k \cdot {\bbox S}_l) +
({\bbox S}_i \cdot {\bbox S}_l)({\bbox S}_k \cdot {\bbox S}_j) \right. \nonumber \\
& &
- \left. ({\bbox S}_i \cdot {\bbox S}_k)({\bbox S}_j \cdot {\bbox S}_l)\right\},
\end{eqnarray}
%%%%%%%%%%%%%%%%%%%%%%%%%%%%%%%%%%%%%%%%%%%%%%%%%%%%%%%%%%%%%%%%%%%%%%%%%
where %{\bf S}$_i$ are spin-1/2 operators and
$J$, $J^{\prime}$ and $J^{\prime\prime}$ are the first-,  second- and
third-nearest-neighbor magnetic exchanges where the paths are illustrated in
Fig.\ \ref{fig1}A. $J_c$ is the ring exchange interaction coupling four spins
(labelled clockwise) at  the corners of a square plaquette. Each spin coupling
is counted once in Eq. (\ref{eq_ham}). Using classical (large-$S$) linear
spin-wave theory the dispersion relation is \cite{MacDonald90,Chubukov92}
$\omega_{\bbox Q}$=$2Z_c({\bbox Q})\sqrt{A_{\bbox Q}^2-B_{\bbox Q}^2}$,
$A_{\bbox Q}$=$J-J_c/2-(J^{\prime}-J_c/4)(1-\nu_h\nu_k)-
J^{\prime\prime}\left[1-(\nu_{2h}+\nu_{2k})/2\right]$,
$B_{\bbox Q}$=$(J-J_c/2)(\nu_h+\nu_k)/2 $, $\nu_x=\cos(2\pi x)$ and
$Z_c({\bbox Q})$ is a renormalization factor \cite{Singh89} that includes the
effect of quantum fluctuations. Within linear spin-wave theory all three
higher-order spin couplings ($J^{\prime},J^{\prime\prime}$ and $J_c$) have
similar effects on the dispersion relation and intensity dependence,
therefore they cannot be determined independently from the data without
additional constraints. We first assume that only $J$ and $J^{\prime}$
are significant as in \cite{Coldea00}, i.e. $J^{\prime\prime}$=$J_c$=0.
The solid lines in Fig.\ \ref{fig2} are fits to a one-magnon cross-section
and Fig.\ \ref{fig3} shows fits to the extracted dispersion relation and
spin-wave intensity. As can be seen in the figures, the model provides an
excellent description of both the spin-wave energies and intensities. The
extracted nearest-neighbor exchange $J$=111.8$\pm$4 meV is antiferromagnetic,
while the next-nearest-neighbor exchange $J^{\prime}$=-11.4$\pm$3 meV across
the diagonal is {\em ferromagnetic}. A wavevector-independent quantum
renormalization factor \cite{Singh89} $Z_c$=1.18 was used in converting
spin-wave energies into exchange couplings. The zone-boundary dispersion
becomes more pronounced upon cooling as shown in Fig.\ \ref{fig3}A and
the dispersion at $T$=10 K can be described by the couplings
$J$=104.1$\pm$4 meV and $J^{\prime}$=-18$\pm$3 meV.
%%%%%%%%%%%%%%%%%%%%%%%%%%%%%%%%%%%%%%%%%%%%%%%%%%%%%%%%%%%%%%%%%%%%%%%%%%%%%%%%%%
%%%%%%%%%%%%%%%%%%%%%%%%%%%%%%%%%%%%%%%%%%%%%%%%%%%%%%%%%%%%%%%%%%%%%%%%%%%%%%%%%%
%%%%%%%%%%%%%%%%%%%%%%%%%%%%%%%%%%%%%%%%%%%%%%%%%%%%%%%%%%%%%%%%%%%%%%%%%%%%%%%%%%
%%%%%%%%%%%%%%%%%%%%%%%%%%%%%%%%%%%%%%%%%%%%%%%%%%%%%%%%%%%%%%%%%%%%%%%%%%%%%%%%%%

A ferromagnetic $J^{\prime}$ contradicts theoretical predictions
\cite{Annett89}, which give an antiferromagnetic superexchange
$J^{\prime}$. Wavevector-dependent quantum corrections
\cite{Canali92} to the spin-wave energies can also lead to a
dispersion along the zone boundary even if $J^{\prime}=0$, but
with sign opposite to our result. Another problem with a
ferromagnetic $J^{\prime}$ comes from measurements on
Sr$_2$Cu$_3$O$_4$Cl$_2$ \cite{Kim99}. This material contains a
similar exchange path between Cu$^{2+}$ ions to that corresponding
to $J^{\prime}$ in La$_2$CuO$_4$ and analysis of the measured
spin-wave dispersion leads to an antiferromagnetic exchange
coupling for this path \cite{Kim99}.

While we can cannot definitively rule out a ferromagnetic
$J^{\prime}$ we can obtain a natural description of the data in
terms of a one-band Hubbard model \cite{Hubbard63}, an expansion
of which yields the spin Hamiltonian in Eq. (\ref{eq_ham}) where
the higher-order exchange terms arise from the coherent motion of
electrons beyond nearest-neighbor sites
\cite{Takahashi77,Roger89,MacDonald90}. The Hubbard Hamiltonian
has been widely used as a starting point for theories of the
cuprates and is given by
%%%%%%%%%%%%%%%%%%%%%%%%%%%%%%%%%%%%%%%%%%%%%%%%%%%%%%%%%%%%%%%%%%%%%%%%%
\begin{equation}
{\cal H}=- t \sum_{{\langle i,j \rangle},\sigma=\uparrow,\downarrow}
\left( c^{\dag}_{i\sigma}c_{j\sigma} + \text{H.c.} \right)
+ U \sum_{i} n_{i\uparrow}n_{i\downarrow},
\label{eq_hubbard}
\end{equation}
%%%%%%%%%%%%%%%%%%%%%%%%%%%%%%%%%%%%%%%%%%%%%%%%%%%%%%%%%%%%%%%%%%%%%%%%%
where $\langle i,j \rangle$ stands for pairs of nearest-neighbors counted once.
Eq. (\ref{eq_hubbard}) has two contributions: the first is the kinetic term
characterized by a hopping energy $t$ between nearest-neighbor Cu sites and the
second the potential energy term with $U$ being the penalty for double occupancy
on a given site. At half filling, the case for La$_2$CuO$_4$, there is one electron
per site and for $t/U \rightarrow 0$, charge fluctuations are entirely suppressed
in the ground state. The remaining degrees of freedom are the spins of the electrons
localized at each site. For small but non-zero $t/U$, the spins interact via a
series of exchange terms, as in Eq.(\ref{eq_ham}), due to coherent electron motion
touching progressively larger numbers of sites. If the perturbation series is
expanded to order $t^4$ (i.e. 4 hops) one regains the Hamiltonian (\ref{eq_ham})
with the exchange constants $J=4t^2/U-24t^4/U^3$, $J_c=80t^4/U^3$ and
$J^{\prime}=J^{\prime\prime}=4t^4/U^3$ \cite{Takahashi77,Roger89,MacDonald90}.
We again fitted the dispersion and
intensities of the spin-wave excitations using these expressions for the exchange
constants and linear spin wave theory. The fits are indistinguishable from those
for variable $J$ and $J^{\prime}$. Again assuming \cite{Zc} $Z_c$=1.18, we obtained
$t$=0.33$\pm$0.02 eV and $U$=2.9$\pm$0.4 eV ($T$=295 K), in agreement with $t$
and $U$ determined from photoemission \cite{Kim98} and optical spectroscopy
\cite{Schuttler92}. The corresponding exchange values are $J$=138.3$\pm$4 meV,
$J_c$=38$\pm$8 meV and $J^{\prime}$=$J^{\prime\prime}$=$J_c/20$=2$\pm$0.5 meV
(the parameters at $T$=10 K are $t$=0.30$\pm$0.02 eV, $U$=2.2$\pm$0.4 eV,
$J$=146.3$\pm$4 meV and $J_c$=61$\pm$8 meV). Using these values, the higher-order
interactions amount to $\sim$11\% ($T$=295 K) of the total magnetic energy
2($J$-$J_c$/4-$J^{\prime}$-$J^{\prime\prime}$) required to reverse one spin
on a fully-aligned N\`{e}el phase.

Many results on oxides of copper fall into place when cyclic exchange of the size extracted
from our experiments is taken into account. First, the relative magnitude of the cyclic
exchange $J_c/J$=0.27$\pm0.06$ at $T$=295 K (0.41$\pm$0.07 at $T$=10 K) is similar to the
ratio of 0.30 estimated from numerical simulations \cite{Schmidt90} on finite clusters
taken from the Cu-O square lattice.
Second, magnetic Raman scattering \cite{Lyons89,Sugai90} and infrared absorption experiments
\cite{Lorenzana99,Perkins93} show an unusual broadening towards higher energies that cannot
be accounted for by a simple (quadratic) Heisenberg Hamiltonian, but can be attributed
\cite{Sugai90,Lorenzana99} to a cyclic term. Finally, in the related compound
Sr$_{14}$Cu$_{24}$O$_{41}$, which has square plaquettes stacked to form a ladder, the
exchange constants corresponding to the nearly-equal-length rungs and legs of the ladder
are 130 meV and 72 meV respectively \cite{Eccleston98} when no cyclic exchange is included.
The inclusion of a ring exchange term \cite{Brehmer99} $J_c$=34 meV allows the rungs and
legs of the ladder to have similar exchange constants of 121 meV.

We have used a new high-wavevector-resolution epithermal-neutron scattering technique
to discover that interactions beyond those coupling nearest-neighbor Cu$^{2+}$ ions
are needed to account for the magnetism of La$_2$CuO$_4$. The observed further neighbor
couplings may be explained by a four-spin cyclic interaction, which arises because
the large orbital hybridization in the CuO$_2$ planes provides an exchange path to
include all four spins at the corners of elementary Cu$_4$O$_4$ plaquettes. Thus,
La$_2$CuO$_4$ joins the nuclear magnet $^3$He \cite{Roger83} as a system where
there is good evidence for substantial ring exchange. Ring exchange occurs even
for very simple models, such as the single band Hubbard model, which contains
only hopping ($t$) and on-site Coulomb ($U$) terms. We have  determined $t$ and $U$
using only knowledge - in the form of our spin wave dispersion relation - of
charge-neutral excitations and find values in excellent agreement with those
obtained via charge-sensitive spectroscopies as well as numerical work on finite
clusters. Thus, our results demonstrate that a one-band Hubbard model is an
excellent starting point \cite{Anderson87} for describing the magnetic
interactions in the cuprates, and that even when considering relatively low
energy spin excitations, charge fluctuations involving double occupancy must
be taken into account. In addition, the scale of the cyclic exchange
interactions, which are comparable to pairing energies in the high-Tc
materials, implies that they themselves or related electronic currents
\cite{Hsu91} might be important for superconductivity in the doped cuprates.

We are grateful to Ris{\o} National Laboratory for help in preparing
this experiment and to J.F. Annett, H.-B. Braun, A.V. Chubukov,
B. Roessli, H.M. R{\o}nnow, G. Sawatzky, Z-X. Shen and R.R.P. Singh
for very helpful discussions. ORNL is managed for the US DOE by
UT-Battelle, LLC, under contract DE-AC05-00OR22725.


\begin{references}
\bibitem{intensities}
In extracting spin-wave intensities we used the
Cu$^{2+}$ 3$d_{x^2-y^2}$ magnetic form factor
and assumed a spherical distribution of ordered
N\`{e}el domains as in \protect\cite{Hayden91}.
\bibitem{Zc}
Using a wavevector-dependent $Z_c({\bbox Q})$ \cite{Canali92} is
expected to modify slightly the extracted values of all exchange
parameters leading to a larger $J_c$. Accurate determinations
require extending the calculation of $Z_c({\bbox Q})$ for the
nearest-neighbor model \cite{Canali92} to the case of finite
higher-order terms as in Eq. (\ref{eq_ham}).
\end{references}
\end{document}